\documentclass[aps,amssymb,reprint,onecolumn,superscriptaddressgroupaddress]{revtex4-1}
\usepackage{amsmath}
\usepackage{hyperref}
\usepackage[english]{babel}
\usepackage[utf8]{inputenc}
\usepackage{graphicx}
\usepackage{braket}
\usepackage{float}
\usepackage{lineno}
\usepackage{babel}
\usepackage{setspace}
\usepackage{subfloat}
\usepackage{xcolor}
\usepackage{datetime}
\usepackage{fancyhdr}
\usepackage{graphicx}
\usepackage{amsmath}
\usepackage{breqn}
\usepackage{setspace}
\usepackage{natbib}
\usepackage{array}
\usepackage{multirow}
\usepackage{longtable}

\begin{document}
\title{On the Order Parameter of the Continuous Phase Transition in the Classical and Quantum Mechanical limits}
	
\author{C. A. M. dos Santos$^1$, F. S. Oliveira$^1$, M. S. da Luz$^2$, and J. J. Neumeier$^3$}

\affiliation{$^1$ Universidade de São Paulo $,$ Escola de Engenharia de Lorena $,$ Lorena - SP $,$ 12602-810 $,$ Brazil}
\affiliation{$^2$Universidade Federal do Triângulo Mineiro$,$ Instituto de Ciências Tecnológicas e Exatas$,$  Uberaba - MG$,$ 38025-180$,$ Brazil}
\affiliation{$^3$Montana State University$,$ Physics Department$,$ Bozeman - MT$,$ 59717-3840$,$ USA}

\begin{abstract}
				
The mean field theory is revisited in the classical and quantum mechanical limits. Taking into account the boundary conditions at the phase transition and the third law of the thermodynamics the physical properties of the ordered and disordered phases were reported. The equation for the order parameter predicts the occurrence of a saturation of $\Psi^2$ = 1 near $\Theta_S$, the temperature below the quantum mechanical ground state is reached. The theoretical predictions are also compared with high resolution thermal expansion data of SrTiO$_{\text{3}}$ monocrystalline samples and other some previous results. An excellent agreement has been found suggesting a universal behavior of the theoretical model to describe continuous structural phase transitions.

		
		
	\end{abstract}

\maketitle
	
	\section{Introduction}
	\label{intro}
	Mean field theory, first developed by Landau \cite{landau1937,landau1959,mnyukh2013}, has successfully described most of the continuous phase transitions, such as structural distortions \cite{bismayer1986a} , magnetic \cite{cracknell1976}, and superconducting transitions \cite{fabrizio2006}, by introducing an order parameter ($\Psi$) which describes many physical properties based upon the fraction of both order and disordered phases coexisting in a given temperature below the critical temperature of the phase transition \cite{mnyukh2013}.
	
	This theory is better applied near the phase transition ($T \sim T_{\text{C}}$), where the density of the ordered phase, given by $\Psi^2$ is small, because the free energy can be computed by a power series of $\Psi$. The solution to minimize the free energy near $T_{\text{C}}$ provides $\Psi \propto$ $\left(T_{\text{C}} - T\right)^{\beta}$, with $\beta$ between 0.25 and 0.50 \cite{sato1985,muller1971}. Some authors have recognized this as the classical limit of the mean field theory \cite{brush1967,onsager1944}.
	
	On the other hand, describing the physical properties at low temperature limit ($T \ll  T_{\text{C}}$) is a challenge since the density of the ordered phase is high ($\Psi^2 \sim 1$) and free energy cannot be expressed by a mathematical series \cite{hayward1999}. This is the quantum limit in which physical properties must reach saturations due to a quantum mechanical ground state \cite{odonnell1991,marques2005,kok2015}.
	
	One of the most successful theoretical description which takes into account the saturation of order parameter has to do with Thomas, Salje and collaborators \cite{salje1991a,salje1991b,salje1992,hayward1999,thomas1971}, who have included a harmonic oscillation term in the free energy due to the soft phonon modes related to the continuous displacive phase transitions \cite{venkataraman1979,cowley2012,bussmann2007,carpenter1998}. The model was successfully applied to described several physical properties of the many materials \cite{salje1992,salje1992a,hayward1999,muller1968,bismayer1986a,bismayer1986b,salje1991a,thomas1971,venkataraman1979,bussmann2007,carpenter1998} and seems to hold an universal behavior for this type of structural phase transition (see for instance figure 1 in reference \cite{salje1991b}).  
	
	Despite the successful of this model, our recent results on high resolution thermal expansion measurements (HRTE) \cite{neumeier2008}, which has relative resolution 100 to 1000 times better than diffractometric techniques \citep{okazaki1973} as well as thousands of data points in each measurement, performed in SrTiO$_{\text{3}}$ single crystals \cite{oliveira2021}, have brought some important insights regarding to the order parameter saturation, especially due to the saturation of the volumetric thermal expansion at low temperatures, which must respect the third law of thermodynamics, i. e. the thermal expansion coefficient ($\Omega = 1/V_{\text{C}}$ $ d \left( \Delta V \right)/ dT $) must be zero as the temperature approaches absolute zero.      
	
	Thus, we have revisited the theoretical model by Salje \textit{et al.} \cite{salje1991b} in order to carefully take into account the boundary conditions at the phase transition ($T = T_{\text{C}}$), which should respect the continuity of the free energy ($G$), volume ($V$), entropy ($S$), and energy ($U$) of the ordered and disordered phases, and, at zero temperature, in which $S$ and $\Omega$ must be equal to zero in order to attend the third law of the thermodynamics \cite{landsberg1956,levy2012}. In addition, the equations for the physical properties in the classical limit should be recovered when the characteristic temperature that holds the ground state in quantum mechanical limit vanishes. 
	
	HRTE measurements performed in SrTiO$_{\text{3}}$ single crystal shown unambiguously quadratic temperature dependences in a large temperature interval below the phase transition. The cubic to tetragonal structural transition in this compound has been well described by the model reported here. We found a direct experimental evidence that the thermal expansion coefficient is the best physical property to describe the order parameter of this transition.
	
	\section{Classical model}
Taking the classical mean field theory by Landau \cite{landau1937,landau1959,mnyukh2013} for a continuous phase transition, the Gibbs free energy is generically given by
\begin{equation}
	G = G_{\text{D}} + a\left(T_{\text{C}} - T \right)\Psi^2+b\Psi^4 + ...
	\label{G_classical}
\end{equation}
where $a$ and $b$ are constants, and $G_{\text{D}}$ is the Gibbs free energy of the disordered phase, when $\Psi$ = 0.
Keeping only the first three terms of the series, the equilibrium order parameter can be obtained by
\begin{equation}
	\frac{dG}{d(\Psi^2)} = 0,
	\label{G_psi}
\end{equation}
which implies
\begin{equation}
	\Psi^2 = 0,
	\label{psi0}
\end{equation}
for disordered phase (D) at $T > T_{\text{C}}$ and,
\begin{equation}
	\Psi^2 = -\frac{a\left(T_{\text{C}} - T \right)}{2b},
	\label{psi2_ab}
\end{equation}
which describes the order parameter and the density of the ordered phase (O) at $T \leq T_{\text{C}}$.
Inserting equation \ref{psi2_ab} into \ref{G_classical} provides
\begin{equation}
	G = G_{\text{D}} - \frac{a^2}{4b}\left(T_{\text{C}} - T \right)^{2}.
	\label{G_GD}
\end{equation}
Taking into account only the effects of entropy and volume in a structural phase transition, the Gibbs free energy is a function of the temperature and pressure, $G = G(T,P)$, that implies
\begin{equation}
	dG = -SdT + VdP
	\label{dG}
\end{equation}
in which
\begin{equation}
	-S = \left(\frac{\partial G}{\partial T}\right)_{P},
	\label{S}
\end{equation}
and
\begin{equation}
	V = \left(\frac{\partial G}{\partial P}\right)_{T}.
	\label{SS}
\end{equation}

Comparing equations \ref{G_GD} and \ref{dG}, and remembering that the entropy of the disordered phase is assumed to be temperature independent, which is given by
\begin{equation}
	S_{\text{D}} = -\left(\frac{\partial G_{\text{D}}}{\partial T}\right)_P,
	\label{S_cubico}
\end{equation}
one can find
\begin{equation}
	G_{\text{D}} = G^0 - S_{\text{D}} T,
	\label{GD}
\end{equation}
where $G^0$ is a constant which defines a reference for the free energy.

Furthermore, taking into account the boundary conditions $G = G_{\text{D}} = G_{\text{O}}$, $S = S_{\text{D}} = S_{\text{O}} = S_{\text{C}}$, and $\Psi^2 = 0$ at the phase transition ($T = T_{\text{C}}$), and $S = 0$ and $\Psi^2 = 1$ at $T = 0$, due to third law of thermodynamics, it is possible to show that
\begin{equation}
	a = S_{\text{C}},
	\label{alpha}
\end{equation}
\begin{equation}
	b = \frac{1}{2}S_{\text{C}} T_{\text{C}},
	\label{b1}
\end{equation}
\begin{equation}
	\Psi^2 = \frac{\left(T_{\text{C}}-T\right)}{T_{\text{C}}}=1 - \frac{T}{T_{\text{C}}},
	\label{psi2}
\end{equation}
and
\begin{equation}
	G = G^0 - S_{\text{C}}T - S_{\text{C}}\left(T_{\text{C}} - T \right)\Psi^2 + \frac{1}{2}S_{\text{C}} T_{\text{C}} \Psi^4,
	\label{GPsi}
\end{equation}
or
\begin{equation}
	G = G^0 - S_{\text{C}}T - \frac{1}{2}\frac{S_{\text{C}}}{T_{\text{C}}}\left(T_{\text{C}} - T\right)^2.
	\label{GTc}
\end{equation}

Taking the derivative of equation \ref{GPsi} with regard to temperature, it is possible to find that
\begin{equation}
	S = S_{\text{C}} \left(1 - \Psi ^2\right) = S_{\text{C}} \frac{T}{T_{\text{C}}}.
	\label{GTc1}
\end{equation}

Furthermore, such as $V$ and $S$ are independent variables, one can use the relation
\begin{equation}
	-\left(\frac{\partial S}{\partial P}\right)_T = \left(\frac{\partial V}{\partial T}\right)_P = \Omega,
	\label{thermoalfa}
\end{equation}
to demonstrate that
\begin{equation}
	-\left(\frac{\partial S}{\partial P}\right)_T = -\left(\frac{\partial S_{\text{C}}}{\partial P}\right)_T \left(1 - \Psi^2\right) - S_{\text{C}}\left(\frac{\partial \left(\Psi ^2\right)}{\partial P}\right)_T,
	\label{thermoPsi}
\end{equation}
or
\begin{equation}
	\Omega = \Omega_{\text{D}}\left(1-\Psi^2\right) + S_{\text{C}} \frac{T}{T_{\text{C}}^2} \left(\frac{d T_{\text{C}}}{d P}\right),
	\label{alfaPsi}
\end{equation}
where $T_{\text{C}}$ holds all the pressure dependence of $\Psi^2$ in the equation \ref{thermoPsi} and $dT_{\text{C}}/dP$ measures the pressure dependence of the critical temperature.

But at $T = T_{\text{C}}$, $\Psi^2 = 0$, which implies that the thermal expansion coefficient of the ordered phase ($\Omega_{\text{O}}$) is different than that of the disordered phase ($\Omega_{\text{D}}$) due to the lambda-type jump at the transition temperature. Thus, from equation \ref{alfaPsi} one can write
\begin{equation}
	\Omega_{\text{D}} = \Omega_{\text{O}} - S_{\text{C}} \frac{1}{T_{\text{C}}} \left(\frac{d T_{\text{C}}}{d P}\right),
	\label{alfaD}
\end{equation}
which put back into equation \ref{alfaPsi}, remembering that $\left(1 - \Psi^2\right)= T/T_{\text{C}}$ from equation \ref{psi2}, leads to 
\begin{equation}
	\Omega = \Omega_{\text{O}} \left(1-\Psi^2\right).
	\label{alfaOP}
\end{equation}

Thus, $\Psi^2$ can be described as a function of the fundamental thermodynamic properties $T$, $S$, or $\Omega$ as
\begin{equation}
	\Psi^2 = 1 - \frac{T}{T_{\text{C}}},
	\label{PsiT_Tc}
\end{equation}
\begin{equation}
	\Psi^2 = 1 - \frac{S}{S_{\text{C}}},
	\label{PsiSSc}
\end{equation}
and,
\begin{equation}
	\Psi^2 = 1 - \frac{\Omega}{\Omega_{\text{O}}},
	\label{Psialfa_alfac}
\end{equation}   
which predict linear dependencies of $S$ and $\Omega$ as a function of the temperature.
	
	\section{Quantum mechanical model}
	
Regarding to the saturation of order parameter at low temperature Thomas, Salje, and other coworkers \cite{salje1991a,salje1991b,salje1992,hayward1999,thomas1971} have proposed a modification of the free energy to take into account quantum mechanical aspects, especially the harmonic oscillations due to soft modes, which are developed below the critical temperature of the structural phase transition. The free energy given by equation \ref{G_classical} from classical limit can be rewritten in the following form related to the quantum mechanical limit
\begin{dmath}
	\label{Gmech}
	G = G_{\text{D}} + \frac{a}{2} \Theta_{\text{S}}\left[\coth{\left(\Theta_{\text{S}} /T\right)} - \coth{\left(\Theta_{\text{S}} /T_{\text{C}} \right)}\right]\Psi^2+\frac{b}{4}\Psi^4.
\end{dmath}

As far as we know, this equation has appeared for the first time in the report by Salje \textit{et al.} in 1991 (see equation 37 in reference \cite{salje1991b}). They have applied it to describe the behavior of many displacive transitions in several compounds. The $\Theta_{\text{S}}$ measures a temperature in which ground state in quantum mechanical limit becomes relevant.

After an extensive mathematical work using similar procedure and the same boundary conditions at $T_{\text{C}}$ and at zero temperature to find the equations for the classical limit, we were able to find the physical parameters $a$, $b$, and $G_{\text{D}}$ of equation \ref{Gmech}, and thermodynamic properties of the continuous phase transition in the quantum mechanical limit (see Appendix). The Gibbs free energy can be rewritten as  

\begin{dmath}
	\label{GibbsMech}
	G = G^0 - S_{\text{C}}\Theta_{\text{S}}\coth{\left(\Theta_{\text{S}} / T \right)} - S_{\text{C}}\Theta_{\text{S}}\left[\coth{\left(\Theta_{\text{S}} / T_{\text{C}} \right)}
	- \coth{\left(\Theta_{\text{S}} / T \right)} \right]\Psi^2 + \frac{1}{2} S_{\text{C}} \Theta_{\text{S}}\left[\coth{\left(\Theta_{\text{S}} /T_{\text{C}}\right)}-1\right]\Psi^4.
\end{dmath}

The first important difference from the previous reports \cite{salje1991a,salje1991b,salje1992,salje1992a,hayward1999,salje1992,carpenter1998} has to do with the first two terms of equation \ref{Gmech}, which are related to free energy ($G_{\text{D}}$) of the disordered phase, that has a temperature dependence more complicated than in the classical limit ($G_{\text{D}} \propto T$). The discussion afterwards will demonstrate that this term plays an important role in the quantum mechanical description of the total entropy in the low temperature regime. Furthermore, the second relevant observation has to do with the third law of thermodynamics, which requires $S$ = zero at zero temperature, implying a saturation in the order parameter at $\Psi^2 = 1$ as $T$ approaches zero ($T < \Theta_{\text{S}}$). This is an important difference since previous reports \cite{,salje1991b,hayward1999} in which it predicts a saturation of $\Psi^2$ at a fraction of one. This has to do with the pre-factor $b$ in equation \ref{Gmech} which normalizes $\Psi^2(T)$ between zero at $T = T_{\text{C}}$ and 1 at $T \rightarrow 0$. Equation \ref{GibbsMech} allowed us to find the order parameter in the low temperature phase as
\begin{equation}
	\Psi^2 = \frac{\coth{\left(\Theta_{\text{S}} /T_{\text{C}}\right)}-\coth{\left(\Theta_{\text{S}} /T\right)}}{\coth{\left(\Theta_{\text{S}} /T_{\text{C}}\right)}-1}.
	\label{PsiTOP}
\end{equation}

Futhermore, equation \ref{GibbsMech} yields the analytical determination of the thermodynamic properties in the quantum mechanical limit, as shown in table 1 (see details in the Appendix). They are compared with those from the classical limit. Equations for the classical limit are naturally recovered when the quantum mechanical characteristic temperature $\Theta_{\text{S}}$ is vanished (compare first and last columns).

\begin{table*}
	\label{tab_eq}       
	\caption{Some thermodynamic properties in the classical and quantum mechanical limits for the mean field model described in this work. Last column displays the equations for the quantum mechanical limit for $\Theta_{\text{S}}$ much smaller than $T$ and $T_{\text{C}}$.}
	\begin{tabular}{>{\centering\arraybackslash}m{2 cm} >{\centering\arraybackslash}m{3.5 cm} >{\centering\arraybackslash}m{6.5 cm} >{\centering\arraybackslash}m{3.5 cm}}
		\hline
		Parameter &
		Classical limit &
		Quantum mechanical limit &
		
		$T \approx T_{\text{C}} \gg \Theta_{\text{S}}$ \\
		\hline
		
		$G_{\text{D}}$ &
		$G^0 - S_{\text{C}}T$ & 
		$G^0 - S_{\text{C}} \Theta_{\text{S}}\coth{\left(\Theta_{\text{S}}/T\right)}$ &
		
		$G^0 - S_{\text{C}}T$ \\ [8 pt]
		
		$S_{\text{D}}$ &
		
		$S_{\text{C}}$ & 
		
		$S_{\text{C}}\left(\Theta_{\text{S}}/T\right)^2$ csch $^2{\left(\Theta_{\text{S}}/T\right)}$ &
		
		$S_{\text{C}}$ \\ [9 pt]
		
		$G$ & 
		
		$G_{\text{D}} - \frac{S_{\text{C}}}{2}\frac{\left(T_{\text{C}} - T\right)^2}{T_{\text{C}}}$ &
		
		$G_{\text{D}} - \frac{S_{\text{C}} \Theta_{\text{S}}}{2}\frac{\left[\coth{\left(\Theta_{\text{S}} / T_{\text{C}}\right)}-\coth{\left(\Theta_{\text{S}} / T \right)}\right]^2}{\left[\coth{\left(\Theta_{\text{S}} / T_{\text{C}}\right)}-1\right]}$ &
		
		$G_{\text{D}} - \frac{S_{\text{C}}}{2}\frac{\left(T_{\text{C}} - T\right)^2}{\left(T_{\text{C}} - \Theta_{\text{S}}\right)}$ \\ [9 pt]
		
		$\Psi^2$ &
		$\left(T_{\text{C}} - T\right) / T_{\text{C}} $ & $\frac{\left[\coth{\left(\Theta_{\text{S}} / T_{\text{C}}\right)} - \coth{\left(\Theta_{\text{S}} / T\right)}\right]}{ \left[\coth{\left(\Theta_{\text{S}} / T_{\text{C}}\right)} -1 \right]}$ & 
		$\left(T_{\text{C}} - T\right) / \left(T_{\text{C}} - \Theta_{\text{S}} \right)$ \\ [8 pt]
		\hline
	\end{tabular}
	
	\begin{tabular}{>{\centering\arraybackslash}m{2 cm} >{\centering\arraybackslash}m{14.5 cm}}
		
		$S$  & 
		
		$S_{\text{D}}\left(1 - \Psi^2\right) $ \\ [8 pt]
		
		$\Omega$  & 
		
		$\Omega_{\text{O}}\left(1 - \Psi^2\right)$  \\ [8 pt]
		\hline
	\end{tabular}
\end{table*}

\begin{figure}
	\includegraphics[width = 8.5 cm]{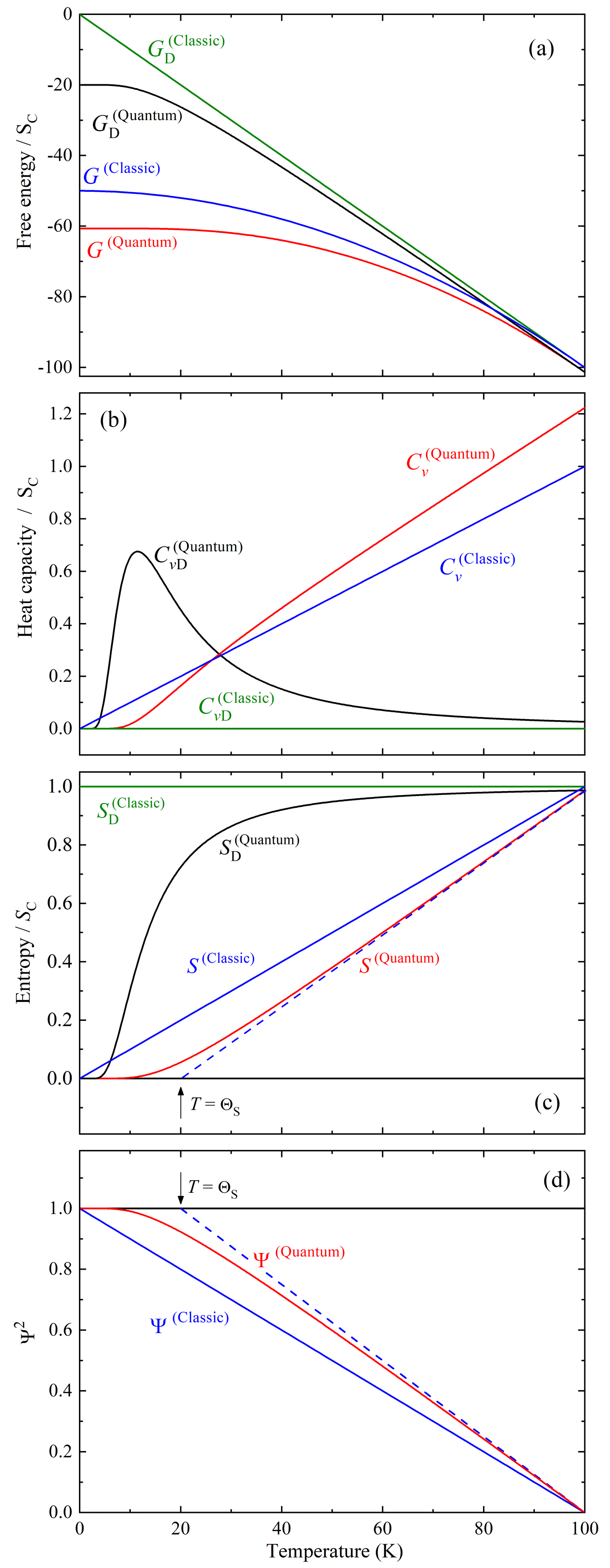} 
	\caption{Predicted behavior in the classical and quantum mechanical limits for the (a) free energy, (b) heat capacity at constant volume, (c) entropy, and (d) order parameter. The curves were plotted using $\Theta_{\text{S}}$ = 20 K and $T_{\text{C}}$ = 100 K.}
	\label{fig:1}       
\end{figure}

In order to better understand the equations in this model, in figure \ref{fig:1} are plotted the behavior of the main properties for the quantum mechanical (black and red lines) and classical limits (blue and green lines) using $\Theta_{\text{S}}$ = 20 K and $T_{\text{C}}$ = 100 K. The results compare the behavior of the properties below $T_{\text{C}}$, which are composed by the contribution of both order and disordered phase densities balanced by the order parameter, with those related only by the disordered phase, indicated with subindex D.

In figure \ref{fig:1}(a) are shown the free energy behavior taking $G^0$ = zero for simplicity. Both curves of each limit reach
the same $G$ value at $T = T_{\text{C}}$, since at this point the phase has the same free energy. In addition, the free energy of the ordered phase is lower than in the disordered phase, in both classical and quantum mechanical limits, as expected due to the earlier phase be energetically favourable. 

Figure \ref{fig:1}(b) displays the expected behavior for the heat capacity at constant volume as a function of the temperature. In the classical limit, $C_v \propto T$ in the ordered phase and zero at disordered phase ($S_{\text{D}}$ is supposed to be constant).

In figure \ref{fig:1}(c) are shown the behaviors for the total entropy in the ordered phase ($S$) and the entropy related to the disordered phase ($S_{\text{D}}$). In the classical model, the entropy due to disordered phase is considered temperature independent (green line), and total entropy decreases linearly proportional to the temperature based upon the ratio $T/T_{\text{C}}$, from $T_{\text{C}}$ down to the ground state at absolute zero (blue line). On the other hand, the results for the quantum mechanical regime show temperature dependencies which must be carefully discussed. First of all, $S$ (red line) tends to zero at finite temperature of the order of $\Theta_{\text{S}}$, which is in agreement with the expected by the quantum ground state and the third law of the thermodynamic. Interesting is the behavior of the total entropy, which is almost linear as a function of the temperature in the interval $T/T_{\text{C}}$ = 0.3 to 1, for the $\Theta_{\text{S}}$ and $T_{\text{C}}$ values used in the figure \ref{fig:1}. This observation has to do with the weak dependence of $S_{\text{D}}$ of approximately 10 $\%$ in this temperature interval. This seems to explain why the classical model, $\Psi^2  \propto \left(T_{\text{C}} - T \right)$, has been frequently used to describe quantum mechanical phase transitions (compare equations for order parameter in table 1). In order to clarify that, we also plotted the total entropy in the linear regime (see blue dash line in fig. 1(c)), which is given by $S = S_{\text{C}} \left( T- \Theta_{\text{S}} \right)/\left( T_{\text{C}} - \Theta_{\text{S}} \right)$. Besides of the expected the agreement near $T_{\text{C}}$, interesting is to notice that the extrapolation to $S$ = zero yields $T = \Theta_{\text{S}}$ directly.

Figure \ref{fig:1}(d) shows the behavior of the $\Psi^2$ as a function of the temperature. $\Psi^2$ in the classical limit is linear from zero up to $T_{\text{C}}$, while in the quantum mechanical regime shows a clear saturation at $\Psi^2$ = 1 in temperatures above absolute zero, which is clearly related to the ground state with $S$ = 0. The expected behavior of $\Psi^2$ for $T \gg \Theta_{\text{S}}$ is also shown in figure \ref{fig:1}(d) (see the dashed line). Its extrapolation to $\Psi^2$ = 1 yields directly $\Theta_{\text{S}}$ =  20 K, which agrees with the fact that of a quantum mechanical ground state is reached close to this temperature.

Furthermore, the temperature dependence of $S_{\text{D}}$ is very important for the reduction of the total entropy of the ordered state which reaches the ground state ($S$ = 0) in a thermal energy of the order of $k_B \Theta_{\text{S}}$. Interesting is to note that not only $S$ goes to zero near $\Theta_{\text{S}}$ but also $S_{\text{D}}$. Thus, we can understand $\Theta_{\text{S}}$ as the temperature below which leads the compound to a quantum mechanical ground state making $S$ to vanish faster than in the classical limit ($S$ = 0 only at $T$ = 0).

\begin{figure}
	\includegraphics[width = 8.5 cm]{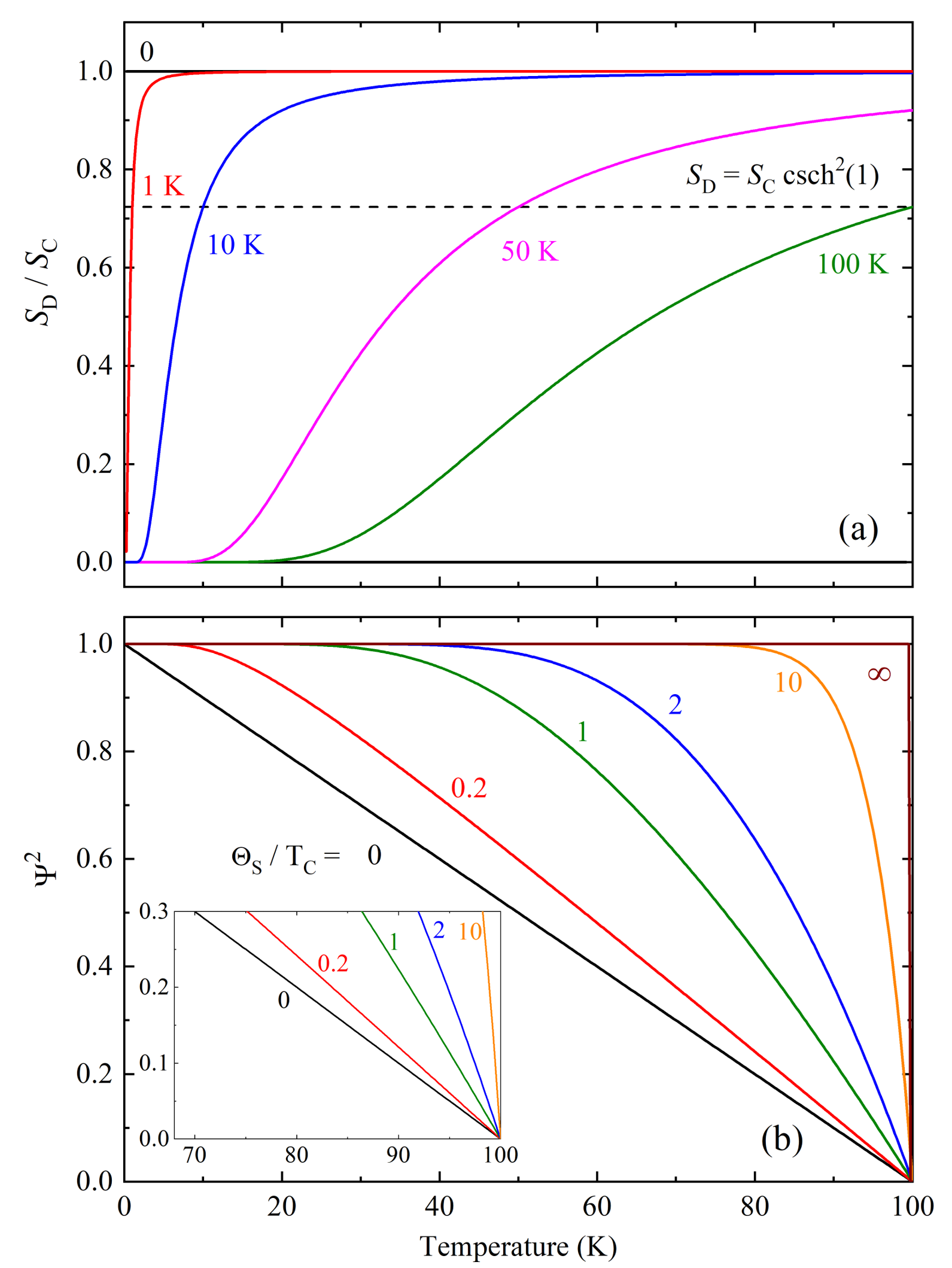} 
	\caption{(a) Temperature dependence of the entropy for the disordered phase for different $\Theta_{\text{S}}$ values. The dashed horizontal line indicates the points where $T = \Theta_{\text{S}}$. When $\Theta_{\text{S}}$ = zero the classical limit of the mean field thery is recovered. In (b) are shown $\Psi^2$ as a function of the temperature for different $\Theta_{\text{S}}/T_{\text{C}}$ values. Inset shows the linear temperature dependence near $T_{\text{C}}$. }
	\label{fig:2}       
\end{figure}

The effect of $\Theta_{\text{S}}$ on $S_{\text{D}}$ is displayed in figure \ref{fig:2}(a) for several different $\Theta_{\text{S}}$ values, remembering that $T_{\text{C}}$ does not play any role on $S_{\text{D}}$. It is possible to observe that the higher $\Theta_{\text{S}}$ the easier the ground state is reached. Furthermore, if one makes $T = \Theta_{\text{S}}$, the equation for $S_{\text{D}}$ given in table 1 leads to $S_{\text{D}} = S_{\text{C}}$ csch$^2(1)=0.724 S_{\text{C}}$, which is shown by the dashed line in figure \ref{fig:2}(a). Interesting is to observe that making $\Theta_{\text{S}}$ = 0, the classical limit is recovered in which $S_{\text{D}}$ = $S_{\text{C}}$ = constant (black line).

In order to evaluate how $\Theta_{\text{S}}$ chances the behavior of the order parameter, in figure \ref{fig:2}(b) are shown some $\Psi^2$ curves as a function of the temperature for different $\Theta_{\text{S}}$ values using a constant $T_{\text{C}}$ = 100 K. It is possible to observe that $\Psi^2$ reaches the ground state at finites temperatures, if $\Theta_{\text{S}} \neq$ 0. Furthermore, the classical behavior is recovered making $\Theta_{\text{S}}$= 0, which represents the linear temperature dependence, $\Psi^2  \propto \left(T_{\text{C}} - T \right)$.

Another important aspect is the shape of the $\Psi^2$ curves, which are extremely dependent of the $\Theta_{\text{S}}/T_{\text{C}}$  ratio. The higher is $\Theta_{\text{S}}/T_{\text{C}}$, the higher is the saturation due to the ground state ($\Psi^2$ = 1 and $S$ = 0). Additionally, one can note that if $\Theta_{\text{S}}/T_{\text{C}}$ ratio tends to infinite, $\Psi^2$ becomes a step-like function (see ref. \cite{venetis2014} and references therein). This behavior reminds a discontinuous (or first order) phase transition, in which the transition from high (disordered) to low (ordered) temperature phase happens abruptly at $T = T_{\text{C}}$ (the origin of this observation will be addressed elsewhere).

Inset of the figure \ref{fig:2}(b) displays the behavior of $\Psi^2$ near $T_{\text{C}}$ for the different $\Theta_{\text{S}}$ values. All the curves show linear temperature dependence given by
\begin{equation}
	\Psi^2 = \left(T_{\text{C}} - T\right) /\left(T_{\text{C}} - \Theta_{\text{S}}\right).
\end{equation}

Due to this linear behavior near $T_{\text{C}}$, probably many authors have used the classical mean field theory to describe phase transitions instead taking into account the quantum mechanical effects.

Although there are many similarities between the model reported here with that reported previously by Salje and coworkers \cite{salje1991b}, but some important differences can be noticed. The most important is, in their results $\Psi^2$ never reaches 1, even at $T$ = 0. We understand this difference because $\Psi^2$ = 1 must happen at $T$ = 0, since a ground state with $S$ = 0 is required due to the third law of the thermodynamics. We have a direct experimental evidence for that using $\Omega$ determined by HTRE experiments performed in SrTiO$_{\text{3}}$ single crystals as shown in the next section.

	\section{Comparison with experiments}
	
\begin{figure}
	\includegraphics[width = 8.5 cm]{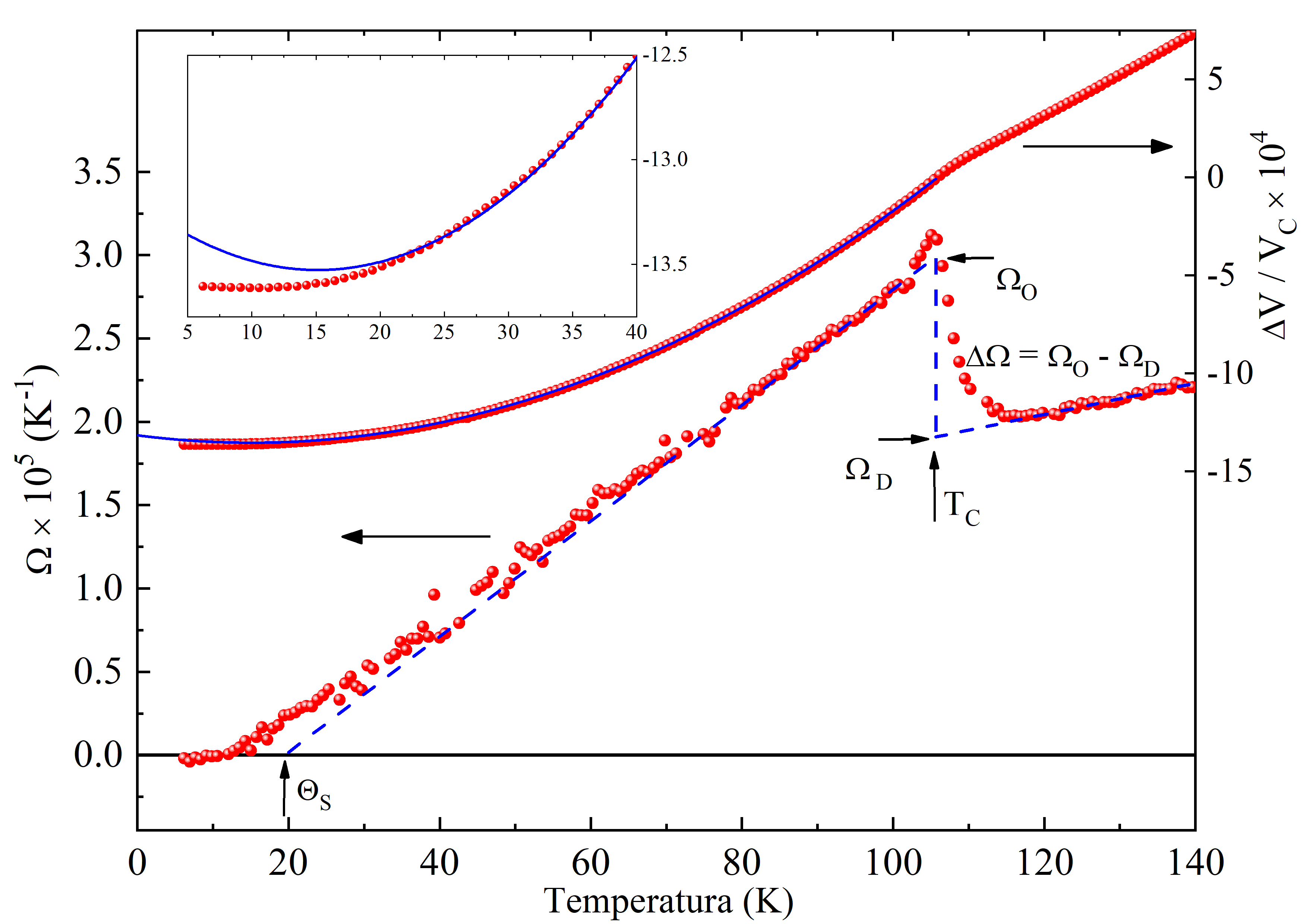} 
	\caption{(Right scale) Volumetric thermal expansion ($\Delta$$V/V$$_{\text{C}}$) measurement of SrTiO$_{\text{3}}$ single crystal (for details see reference \cite{oliveira2021}). A clear quadratic temperature dependence can be observed and is indicated by the fitting displayed by the blue line. Inset displays a magnification at low temperature in which the expected deviation of the quadratic behavior due to the saturation of $\Delta$$V/V$$_{\text{C}}$ is observed. (Left scale) Thermal expansion coefficient determined from $\Omega = d(\Delta$$V/V$$_{\text{C}})/dT$. The linear behavior and the saturation at low temperature can be noticed. $T_{\text{C}}$, $\Omega_{\text{O}}$, $\Omega_{\text{D}}$, and $\Delta \Omega$ are directly determined from linear extrapolation near the phase transition. Only 25$\%$ of the data points are shown in both curves \cite{noise}.}
	\label{fig:3}       
\end{figure}

Recent HRTE measurements have been performed by us in SrTiO$_{\text{3}}$ single crystals \cite{oliveira2021}. Figure \ref{fig:3} shows the temperature behavior of the volumetric thermal expansion ($\Delta V/V_{\text{C}}$) measurement performed using a capacitance quartz cell, which shows a clear quadratic behavior in a large temperature interval below the phase transition temperature (105.65 \,K). This result was also observed in other two oxygen vacancy doped SrTiO$_{\text{3}}$ single crystals. Thanks to HRTE \cite{neumeier2008} which has resolution 100 to 1000 times better than diffraction methods \cite{okazaki1973}, has better precision than metallic cells \cite{liu1997,tsunekawa1984}, and provides thousands of data points in the temperature measurement interval. 

Figure \ref{fig:3} also shows the volumetric thermal expansion coefficient ($\Omega$). A clear linear behavior is observed from $\sim$ 30 K up to near the critical transition temperature. Furthermore, a saturation at low temperature is also clearly noticed, in which $\Omega$ approaches zero at low temperature. Based upon these results, we see a direct connection with the model for the quantum mechanical limit described in section 2. Additionally, as pointed out in several previous works \cite{muller1968,hayward1999}, the rotation angle $\varphi$, which measures the antidistortive angle from the cubic to tetragonal in the transition of the SrTiO$_{\text{3}}$ compound has been directly related as the order parameter. However, our recent results  on HRTE \cite{oliveira2021} suggest that is better to use $\Omega$ as the order parameter instead $\varphi$, since the last one is proportional to $\int \Omega dT$, which is not necessarily zero at $T$ = 0. Thus, in agreement to the model developed in this work $\Omega$, should be related to the $\Psi^2 $, such as shown in table 1 and in Appendix. Hereafter, we discuss the implication of these observations on the HRTE results obtained in SrTiO$_{\text{3}}$.

First of all, the equation for thermal expansion coefficient in the quantum mechanical limit near the critical temperature can be written as
\begin{equation}
	\label{alfa_exp}
	\Omega = \Omega_{\text{O}}\left(\frac{T - \Theta_{\text{S}}}{T_{\text{C}} - \Theta_{\text{S}}}\right).
\end{equation}

Thus, taking the temperature at the peaks as the critical temperature $T_{\text{C}}$ = 105.65 K and making a linear fit (shown by the dashed blue line) yields directly $\Omega_{\text{O}}  = 2.99 \times 10^{-5}$ K $^{-1}$ and $\Theta_{\text{S}}$ = 19.5 K, without much efforts to find the fitting parameters as in previous reports \cite{hayward1999}. Additionally, one must keep in mind that $\Omega$ must be taken subtracting off the background in order to make $\Psi^2$ zero right above $T_{\text{C}}$, as required by the mean field  theory.  

\begin{figure}
	\includegraphics[width = 8.5 cm]{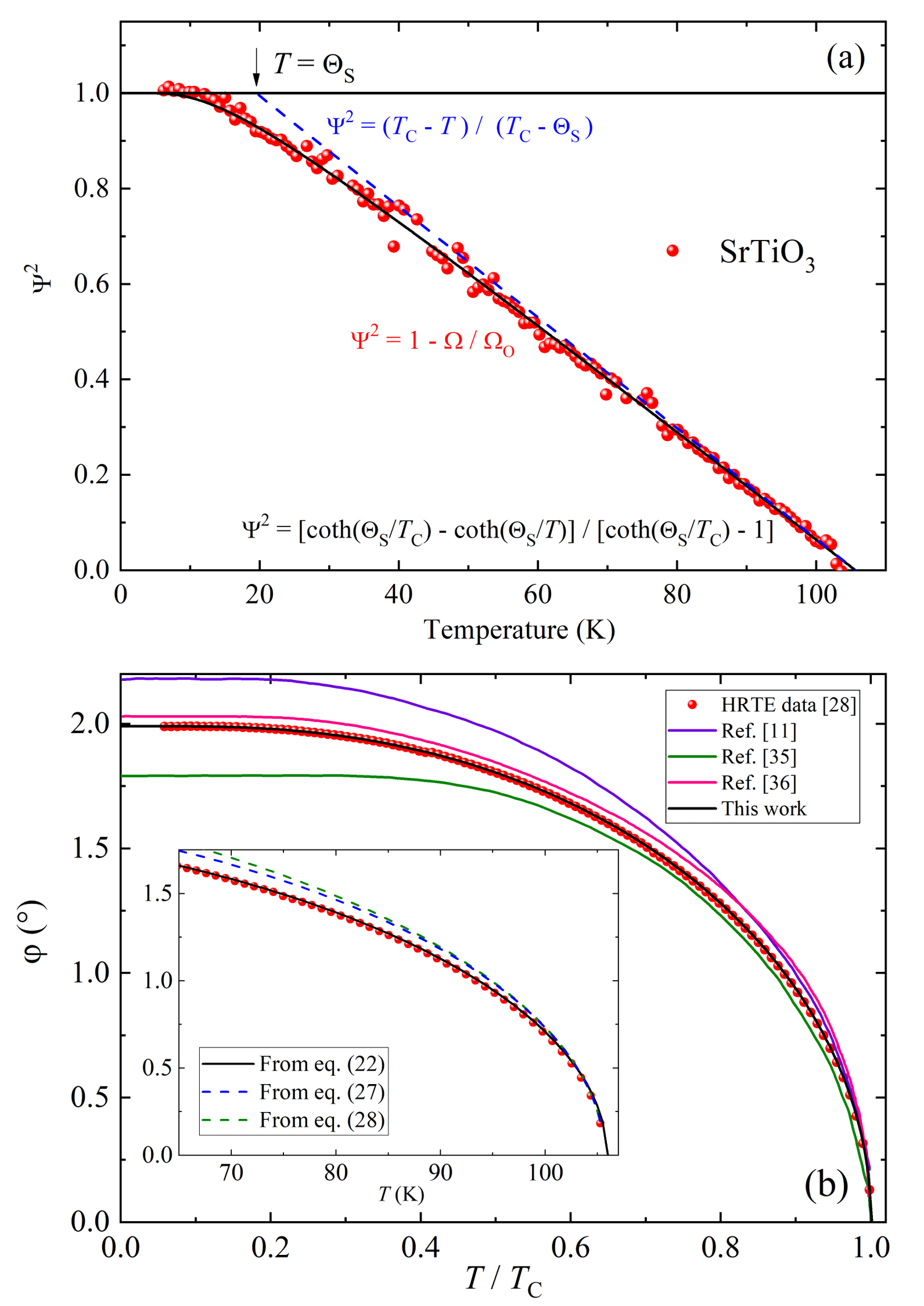} 
	\caption{Comparison of the order parameter determined from HRTE data for SrTiO$_{\text{3}}$ single crystal  with theoretical prediction. $T_{\text{C}}$, $\Theta_{\text{S}}$, and $\Omega_{\text{O}}$ were determined in Fig. \ref{fig:3}. Blue dashed line describes the expected behavior near $T \approx T_{\text{C}}$. In (b) is shown the distortive angle $\varphi$ for the cubic to tetragonal phase transition in SrTiO$_{\text{3}}$ determined from HRTE measurements \cite{oliveira2021} and by numerical integration using the equation for $\Omega$. Other theoretical predictions reported previously are plotted along in order to compare the fittings \cite{pytte1969,feder1970,hayward1999}. Insert compares the experimental data with the theoretical predictions near $T_{\text{C}}$. }
	\label{fig:4}       
\end{figure}

Now, the temperature dependence of the $\Psi^2$ and $\Omega$, given by the correspondent equations in the table 1, can be compared. An excellent agreement between experimental data and theoretical curves in full temperature interval below $T_{\text{C}}$ can be noticed. Saturation near $\Psi^2$ = 1 can be clearly observed, as expected. Additionally, the blue dashed line fits well the behavior near $T_{\text{C}}$ and extrapolates to $T = \Theta_{\text{S}}$ at $\Psi^2$ = 1. This demonstrate that the $\Omega$ instead $\varphi$ is the best order parameter to describe the antidistortive phase transition in the SrTiO$_{\text{3}}$, in agreement with our recent work \cite{oliveira2021}.

In order to show the quality of the agreement between the experimental data and the theoretical description in this work, we compare in figure \ref{fig:4}(b) the results for the angle $\varphi$, obtained from HRTE measurements in SrTiO$_{\text{3}}$ samples, with the theoretical prediction for $\Omega(T)$, using $T_{\text{C}}$, $\Omega_{\text{O}}$, and $\Theta_{\text{S}}$ directly obtained from figure \ref{fig:3}(a) (for more details, see reference \cite{oliveira2021}) with some other theoretical curves reported previously \cite{pytte1969,feder1970,hayward1999,thomas1971}. Although all the theoretical models show fits close to the experimental data, the model proposed here shows the best fit. Furthermore, the previous reports \citep{hayward1999} are based upon fittings which need 4 to 6 parameters. In the present work, the direct determination of parameters from $\Omega(T)$ near the phase transition allows us to find the temperature dependence of $\varphi$ in the full temperature range, which suggest that the model is correct.  

\begin{figure}
	\includegraphics[width = 8.5 cm]{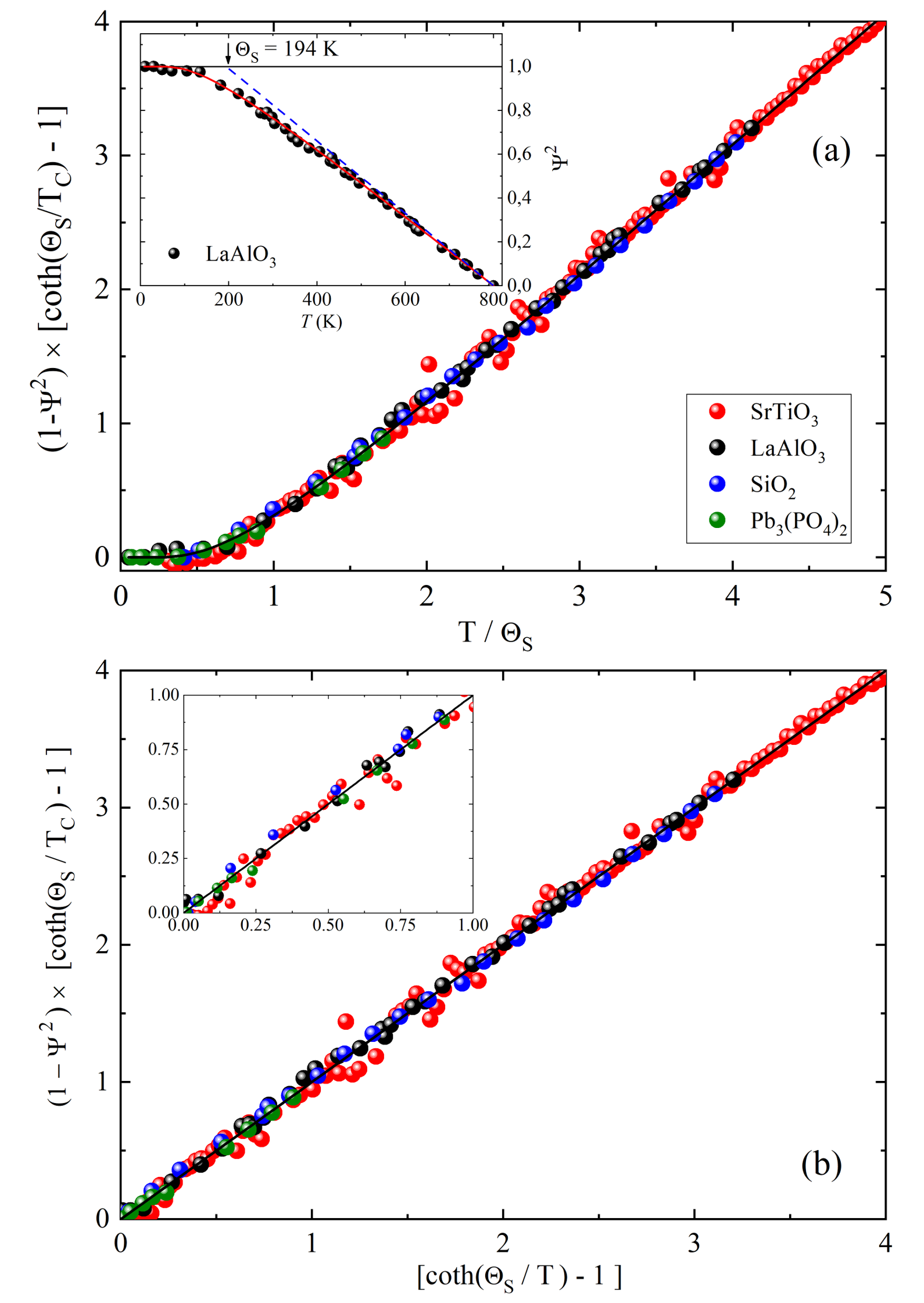} 
	\caption{In (a) is displayed the collapse of the order parameter data previously reported in the references \cite{salje1991b,muller1968,bismayer1986b} using equation \ref{linearpsi}, plotted along with the HRTE data for the SrTiO$_3$ single crystal. Insert shows the way $T_{\text{C}}$ and $\Theta_{\text{S}}$ are determined. They agree with the data reported for LaAlO$_{\text{3}}$ \cite{muller1968}. (b) It is shown the collapse in linear form for the data shown in (a). Inset displays the collapse around the saturation of $\Psi^2$.}  
	\label{fig:5}       
\end{figure}

Finally, we compare the theoretical model for the quantum mechanical limit with some data available on literature, especially reported by Salje and coworkers \cite{salje1991b}. Figure \ref{fig:5} displays the scaling of the data for our SrTiO$_{\text{3}}$ data \cite{oliveira2021}, along with SiO$_{\text{2}}$ \cite{salje1992a}, LaAlO$_{\text{3}}$ \cite{muller1968}, and  Pb$_{\text{3}}$(PO$_{\text{4}}$)$_{\text{2}}$ \cite{bismayer1986b}, all related to the $n$ = 2 in equation 37 reported by Salje \cite{salje1992} (the data available for $n$ = 4 will be addressed elsewhere) based upon the following equation.
\begin{equation}
	\left(1-\Psi^2\right)\left[\coth{\left(\Theta_{\text{S}} / T_{\text{C}}\right)}-1\right] = f\left(\Theta_{\text{S}} / T\right),
	\label{linearpsi}
\end{equation}
where $f\left( \Theta_{\text{S}} / T \right) = \coth{\left(\Theta_{\text{S}} / T\right)} - 1$ and $T/\Theta_{\text{S}}$ is the reduced temperature which measures the ration between thermal energy and quantum mechanical energy that leads the ordered phase to the ground state.

An excellent collapse, shown in figure \ref{fig:5}(a) for the data of all samples and the theoretical prediction displayed by the black lines, are clearly observed, suggesting a universal behavior, despite the definition of the order parameter chosen in the previous reports \cite{muller1968,hayward1999,salje1991b}. The linear behavior displayed in the figure \ref{fig:5}(b) confirms the agreement between the experimental and the quantum mechanical model with the corrections introduced in this works.
	
	\section{Conclusion}
	
Quantum mechanical model for the order parameter has been revisited. Taking into account the boundary conditions, at the continuous phase transition and at the absolute zero temperature, which must obey the third law of thermodynamics, the pre-factor terms of the free energy equation were naturally found. 

Based upon free energy equation, the temperature dependencies of the physical properties related to the order-disordered phase transition were derived. The theoretical model showed that the entropy of the disordered phase plays a very important role in the ordered state, since it has a strong temperature dependence, which reaches a ground state near the characteristic temperature, $\Theta_{\text{S}}$, defined previously by Salje \textit{et al.} \cite{salje1991b}. Furthermore, it also carries on the total entropy to zero near the same temperature.

Interesting is to note that the model predicts that the order parameter is related to one of the three fundamental properties, temperature, entropy, or thermal expansion coefficient. The experimental results on HRTE performed in SrTiO$_{\text{3}}$ single crystals \cite{oliveira2021} provide direct evidence that the volumetric thermal expansion coefficient is the appropriated fundamental physical property to describe the order parameter of the cubic to tetragonal distortive phase transition in this compound, instead the antidistortive angle $\varphi$ \cite{muller1968,hayward1999,salje1991b}. Another evidence that the model works well is the universal collapse of the previous results for the order parameter, both in linear limit ($\Psi^2 \rightarrow 0$) and also at the saturation regime ($\Psi^2 \rightarrow 1$), for the structural continuous phase transitions in other compounds \cite{salje1991b,scott1974}. The fits of these data need only the determination of $T_{\text{C}}$ and $\Theta_{\text{S}}$, in comparison with previous reports, which have to find 4 to 6 fitting parameters \cite{hayward1999}.

Finally, preliminary analyses of other experimental data suggest that the theoretical model reported here can also be applied to other types of continuous phase transitions, such as magnetic and superconducting transitions. In such cases, the energy related to each transition must be added to the entropy term in the free energy equation.

\begin{acknowledgements}
	This work is based upon support by the FAPESP (2009/54001-2 and 2019/12798-3), FAPEMIG (PPM-00559-16), CNPq (308135/2017-2), and CAPES - Finance code 001. Work at Montana State University was conducted with financial support from the US Department of Energy (DOE), Office of Science, Basic Energy Sciences (BES) under Award No. DE-SC0016156.
\end{acknowledgements}
	
	\bibliographystyle{apsrev4-1}
	\bibliography{bjp_bib.bib}

\begin{thebibliography}{37}
\bibitem [{noi()}]{noise}%
  \BibitemOpen
  \href@noop {} {\bibinfo  {journal} {The noise in $\Omega$ curve has to do
  with derivative point by point which is very susceptible to any small
  unexpected change in temperature or capacitance, due to the high sensitivity
  instrument used for the HRTE measurements (see ref. 26).}\ }\BibitemShut
  {NoStop}%
\end{thebibliography}
	
	\begin{center}
		{\bf APPENDIX}
	\end{center}
	
	\renewcommand{\theequation}{A-\arabic{equation}}
	\setcounter{equation}{0}  
	
Taking the quantum mechanical model for the continuous transition predicted by Salje \textit{et al.} \cite{salje1991b}, the free energy is given generically by
\begin{equation}
	G = G_{\text{D}} + a \left[\coth\left(\Theta_{\text{S}}/T\right)-\coth\left(\Theta_{\text{S}}/T_{\text{C}} \right) \right]\Psi^4 + b\Psi^4.
	\label{G_apx}
\end{equation}
Making
\begin{equation}
	\frac{\partial G}{\partial \left(\Psi^2\right)} = 0,
\end{equation}
provides
\begin{equation}
	\Psi^2 = 0,
\end{equation}
for $T > T_{\text{C}}$ (disordered phase), and
\begin{equation}
	\Psi^2 = - \frac{a\left[\coth{\left(\Theta_{\text{S}}/T\right)} - \coth{\left(\Theta_{\text{S}}/T_{\text{C}} \right)}\right]}{2b},
\end{equation}
for $T \le T_{\text{C}}$ (ordered phase).

Taking $\Psi^2 = 1$ for $T$ = 0, implies
\begin{equation}
	\frac{2b}{a} = - \left[1 - \coth{\left(\Theta_{\text{S}}/T_{\text{C}}\right)}\right],
\end{equation}
which is a normalization for $\Psi^2$.

Thus
\begin{equation}
	\Psi^2 = \frac{\left[\coth{\left(\Theta_{\text{S}}/T\right)} - \coth{\left(\Theta_{\text{S}}/T_{\text{C}} \right)}\right]}{\left[\coth{\left(\Theta_{\text{S}}/T_{\text{C}}\right)} - 1\right]},
	\label{PsiNorm}
\end{equation}
implies
\begin{dmath}
	G = G_{\text{D}} - a \left[\coth{\left(\Theta_{\text{S}}/T_{\text{C}}\right)} - \coth{\left(\Theta_{\text{S}}/T\right)} \right] \Psi^2 - a/2\left[1 - \coth{\left(\Theta_{\text{S}}/T_{\text{C}}\right)}\right]\Psi^4,
\end{dmath}
or
\begin{dmath}
	G = G_{\text{D}} - \frac{a}{2}\frac{\left[\coth{\left(\Theta_{\text{S}} / T_{\text{C}}\right)}-\coth{\left(\Theta_{\text{S}} / T \right)}\right]^2}{\coth{\left(\Theta_{\text{S}} / T_{\text{C}}\right)}-1}.
\end{dmath}

Such as $G = G(T,P)$, then
\begin{equation}
	\left(\frac{\partial G}{\partial T} \right)_P = - S,
\end{equation}
or
\begin{equation}
	-S = -S_{\text{D}} + a\Psi^2\left(\Theta_{\text{S}}/T^2\right)\text{csch}^2\left(\Theta_{\text{S}}/T\right),
\end{equation}
which provides the following equation for the entropy of the disordered phase, making $S$ = 0 and $\Psi^2$ =  1 as $T \rightarrow$ 0,
\begin{equation}
	S_{\text{D}} = a \left(\Theta_{\text{S}}/T^2\right)\text{csch}^2\left(\Theta_{\text{S}}/T \right).
	\label{SD_apx}
\end{equation}
Futhermore, when $\Theta_{\text{S}} \rightarrow$ 0, $S_{\text{D}} \rightarrow S_{\text{C}}$, and $a \rightarrow S_{\text{C}}\Theta_{\text{S}}$ due to the classical limit which can be noticed in figure \ref{fig:2}(a). Thus,
\begin{equation}
	S_{\text{D}} = S_{\text{C}}\left(\Theta_{\text{S}}/T\right)^2\text{csch}^2\left(\Theta_{\text{S}}/T \right)
	\label{SD_apx1},
\end{equation}
but taking
\begin{equation}
	\left(\frac{\partial G_{\text{D}}}{\partial T} \right)_P = - S_{\text{D}},
\end{equation}
one can show that
\begin{equation}
	G_{\text{D}} = G^0 - S_{\text{C}} \Theta_{\text{S}} \coth{\left(\Theta_{\text{S}}/T \right)},
	\label{GD_apx}
\end{equation}
where $G^0$ is a reference for the free energy and appears due to the integration constant.

Thus,
\begin{dmath}
	G = G^0 
	- S_{\text{C}}\Theta_{\text{S}}\coth{\left(\Theta_{\text{S}}/T\right)}
	- S_{\text{C}}\Theta_{\text{S}}\left[\coth{\left( \Theta_{\text{S}}/T_{\text{C}} \right)} - \coth{\left(\Theta_{\text{S}}/T \right)} \right]\Psi^2 
	- \frac{S_{\text{C}}\Theta_{\text{S}}}{2}\left[ 1 - \coth{\left(\Theta_{\text{S}} / T_{\text{C}}\right)} \right] \Psi^4.
	\label{GD_salje}
\end{dmath}
which is the equation by Salje \textit{et al} \cite{salje1991b}, with pre-factors determined based upon the boundary conditions at the critical temperature and absolute zero.

Then,
\begin{equation}
	G = G_{\text{D}} - \frac{S_{\text{C}} \Theta_{\text{S}}}{2}\frac{\left[\coth{\left(\Theta_{\text{S}} / T_{\text{C}}\right)}-\coth{\left(\Theta_{\text{S}} / T \right)}\right]^2}{\coth{\left(\Theta_{\text{S}} / T_{\text{C}}\right)}-1},
	\label{G_apx_full}
\end{equation}
which provides
\begin{equation}
	-S = -S_{\text{D}} + S_{\text{C}}\Theta_{\text{S}}\Psi^2\text{csch}^2\left(\Theta_{\text{S}}/T\right)
	\label{S_apx_full}
\end{equation}
or
\begin{equation}
	S = S_{\text{D}} - S_{\text{D}}\Psi^2,
\end{equation}
that leads to
\begin{equation}
	S = S_{\text{D}}\left(1 - \Psi^2\right),
	\label{Spsi_apx}
\end{equation}
which has the same format as the equation for the classical limit, but takes into account the $\Psi^2$ saturation effect near $T = \Theta_{\text{S}}$.

With reagard to thermal expansion, it can be obtained using the following relation for $\Omega$
\begin{equation}
	- \left(\frac{\partial S}{\partial P}\right)_T = - \left(\frac{\partial V}{\partial T}\right)_P = \Omega,
\end{equation}
which provides
\begin{dmath}
	- \left(\frac{\partial S}{\partial P}\right)_T = - \left(\frac{\partial S_{\text{D}}}{\partial P}\right)_T \left(1-\Psi^2\right)-S_{\text{D}}\left(\frac{\partial \left(\Psi^2\right)}{\partial P}\right)_T,
\end{dmath}
where
\begin{dmath}
	\left(\frac{\partial \left(\Psi^2\right)}{\partial P}\right)_T =
	\frac{\Theta_{\text{S}}}{T_{\text{C}}^2} 
	\frac{\left[\coth{\left(\Theta_{\text{S}}/T\right)}-1\right]}{\left[\coth{\left(\Theta_{\text{S}}/T_{\text{C}}\right)}-1\right]^2}
	\text{csch}^2\left(\Theta_{\text{S}} / T_{\text{C}}\right)\left(\frac{d T_{\text{C}}}{d P}\right)_T,
\end{dmath}
and $\Theta_{\text{S}}$ is assumed to be constant with pressure and dependent only of the temperature. Thus, one can write
\begin{dmath}
	\Omega = \Omega_{\text{D}}\left(1-\Psi^2\right)
	+S_{\text{D}}\frac{\Theta_{\text{S}}}{T_{\text{C}}^2}
	\frac{\left[\coth{\left(\Theta_{\text{S}}/T\right)}-1\right]}{\left[\coth{\left(\Theta_{\text{S}}/T_{\text{C}}\right)}-1\right]^2}
	\text{csch}^2\left(\Theta_{\text{S}} / T_{\text{C}}\right)\left(\frac{d T_{\text{C}}}{d P}\right)_T,
	\label{omg_apx}
\end{dmath}
but at $T$ = $T_{\text{C}}$ and $\Omega$ = $\Omega_{\text{O}}$, providing
\begin{dmath}
	\Omega_{\text{D}} = \Omega_{\text{O}}
	- S_{\text{D}}\frac{\Theta_{\text{S}}}{T_{\text{C}}^2}
	\frac{\text{csch}^2\left(\Theta_{\text{S}}/T_{\text{C}}\right)}{\left[\coth{\left(\Theta_{\text{S}}/T\right)}-1\right]}
	\left(\frac{d T_{\text{C}}}{d P}\right)_T,
\end{dmath}
which gives
\begin{dmath}
	\Omega = \Omega_{\text{O}}\left(1-\Psi^2\right)
	- S_{\text{D}}\frac{\Theta_{\text{S}}}{T_{\text{C}}^2}
	\frac{\text{csch}^2\left(\Theta_{\text{S}}/T_{\text{C}}\right)}{\left[\coth{\left(\Theta_{\text{S}}/T\right)}-1\right]}
	\left(\frac{d T_{\text{C}}}{d P}\right)_T
	\left(1-\Psi^2\right)
	+S_{\text{D}}\frac{\Theta_{\text{S}}}{T_{\text{C}}^2}
	\frac{\left[\coth{\left(\Theta_{\text{S}}/T\right)}-1\right]}{\left[\coth{\left(\Theta_{\text{S}}/T_{\text{C}}\right)}-1\right]^2}
	\text{csch}^2\left(\Theta_{\text{S}} / T_{\text{C}}\right)\left(\frac{d T_{\text{C}}}{d P}\right)_T.
	\label{omg_apx1}
\end{dmath}
After an algebraic work it is possible to show that the terms multiplied by $S_{\text{D}}$ cancel each other, remembering that $\left(1 - \Psi^2\right)$ is given by
\begin{equation}
	\left(1-\Psi^2\right) = \frac{\coth{\Theta_{\text{S}}/T}-1}{\coth{\Theta_{\text{S}}/T_{\text{C}}}-1},
\end{equation} 
thus
\begin{equation}
	\Omega = \Omega_{\text{O}}\left(1-\Psi^2\right).
\end{equation}
Finally, based upon the equations for $G_{\text{D}}$, $G$,  $S_{\text{D}}$, and $S$ it is possible to find temperature dependences for heat capacity at constant volume ($C_v$) and the internal energy ($U$) (not shown).

\end{document}